\documentclass[fleqn,12pt,twoside]{article}
\usepackage{espcrc1}
\usepackage{graphicx}
\usepackage[figuresright]{rotating}
\title{Neutrino mean free path in neutron stars}
\author{U. Lombardo\address[LNS]{INFN-LNS,Via S. Sofia 44, I-95123 Catania, Italy}
     \address{Dipartimento di Fisica, Via S. Sofia 64, I-95123 Catania, Italy},
     Caiwan Shen\addressmark[LNS]
     \address{China Institute of Atomic Energy, P.O.Box 275(18), Beijing 102413, China}, 
     N. Van Giai \address{Institut de Physique Nucl\'eaire, F-91406 Orsay, France}, 
     W. Zuo\address{Institute of Modern Physics, Chinese Academy of Science, Lanzhou, China}}
\begin{document}

\maketitle

\begin{abstract}
The Landau parameters of nuclear matter and neutron matter are extracted from the Brueckner 
theory including three-body forces. The dynamical response function to weak neutrino current 
is calculated in terms of the Landau parameters in the RPA limit. Then,
the neutrino mean free path in neutron stars is calculated for different conditions of 
density and temperature. 
\end{abstract} 

\section{INTRODUCTION}
The interaction of neutrinos with baryons has been mostly studied in
connection with the stability of nuclei. But it also plays a crucial role  
in the thermal evolution of supernovae and neutron stars, where the nuclear medium is far from 
stability. A large effort has been 
devoted to study the production of neutrinos via  direct or modified URCA processes
\cite{LATT} and, more recently, via the bremsstrahlung from nucleons in the strong magnetic 
field of neutron stars\cite{SEDR}.  The propagation of neutrinos in neutron matter has 
received less attention\cite{IWA,MARG}.
Being the
experimental information on this topic still insufficient, the theoretical predictions have 
to be developed further, extending the microscopic
description of many-body systems to a wide range of density, temperature and isospin asymmetry. 
This means extending the study of the nuclear equation
of state far away from the saturation point.

In this note we report on the calculation of the response function to
weak interaction coupling in neutron matter. It is based on the 
equation of state (EOS) predicted by the Brueckner theory, including relativistic and finite-size
effects (nucleon resonances). Recently it has made a step
forward as to reproducing the empirical saturation properties of nuclear matter\cite{UMB}. 
The effective NN interaction is extracted from the Brueckner theory and cast in terms of the 
Landau parameters. Then, the response function is calculated in RPA along with the neutrino
mean free path in neutron matter 
for various conditions of neutron density and temperature.

\section{NUCLEAR MATTER EOS FROM THE BRUECKNER THEORY} 
 
The EOS of nuclear matter has been investigated from several microscopic
approaches. Their common feature is that of adopting as $V_{NN}$ interaction  a 
realistic potential, i.e. fitted on the experimental phase shifts of the
nucleon-nucleon scattering in the vacuum. Among them the Brueckner theory has 
proved to be very powerful for its capability of incorporating both relativistic  
and nucleon finite-size effects \cite{GRAN}. This feature is illustrated in
Fig.~1, where a calculation with the interaction Argonne $V_{18}$ is plotted. In 
the first order of the hole-line expansion (BHF approximation), the saturation 
density turns out to be too large, when the NN interaction is just
a two-body force (2BF). Extending the BHF calculation to include a three-body 
force (3BF) effects one improves
the agreement with the empirical saturation properties of nuclear matter.
The origin of 3BF is twofold: the $\overline{N}N$ excitations missing from 
the non relativistic Brueckner theory, amount to a special 3BF as shown
by G.E. Brown et al.\cite{GEB}; low-lying nucleonic excitations as intermediate 
states in the interaction between two nucleons are pure 3BF.
As shown in Fig.1 both effects play a important role for the reproduction
of the empirical saturation properties of nuclear matter.
\begin{figure}
\centering{\includegraphics[width=9cm]{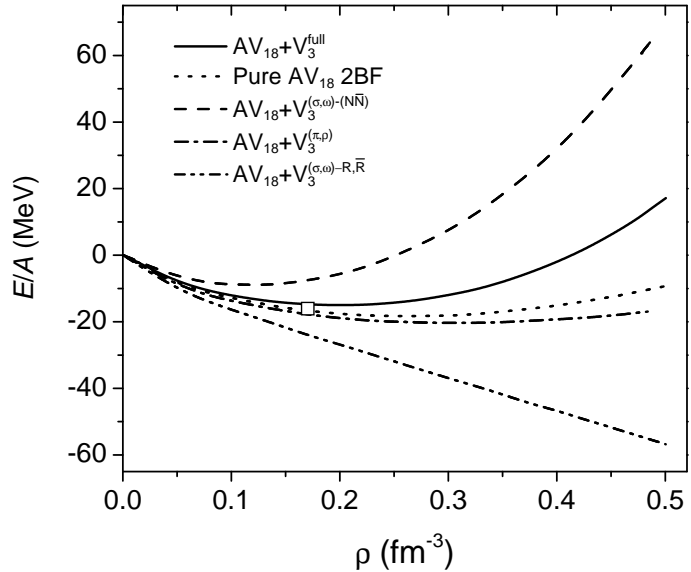}
\vspace{-7mm}
\caption{EOS of symmetric nuclear matter from the Brueckner theory. The effects 
of different components of the 3BF are separately shown. For more
details see Ref.\cite{UMB}. }}
\end{figure}

The predictions of the EOS can be extended far beyond the saturation density 
of cold and symmetric nuclear matter. A variety 
of physical situations can be studied, including neutron-rich  and 
spin-polarized matter at finite temperature (here we limit ourselves to
zero temperature). For such cases there are less physical constraints: 
the isospin-symmetry energy at the saturation point is constrained by the 
empirical value of $\approx 30 MeV$ extracted from the nuclear mass
table: finally, the spin-symmetry energy can be probed by spin-flip
excitations of nuclei.

A set of Brueckner calculations have been performed in different states of spin 
and isospin polarization of nuclear matter, defined by
\begin{equation}
\delta = \frac{A_{\uparrow}-A_{\downarrow}}{A},\quad \beta=\frac{N-Z}{A},
\end{equation} 
respectively. Here $A_{\uparrow}$($A_{\downarrow}$) are the number of nucleons 
with spin polarized upward (downward).
In both cases the Brueckner calculations predict a quadratic law so that the isospin 
and spin symmetry energies describe completely the properties of isospin and spin 
polarized nuclear matter, respectively. In Fig.~2 the isospin symmetry 
energy and in Fig.~3 the spin symmetry energy are plotted as a function of density.
\begin{figure}
\begin{minipage}[t]{74mm}
\includegraphics[angle=0,width=65mm,height=7cm]{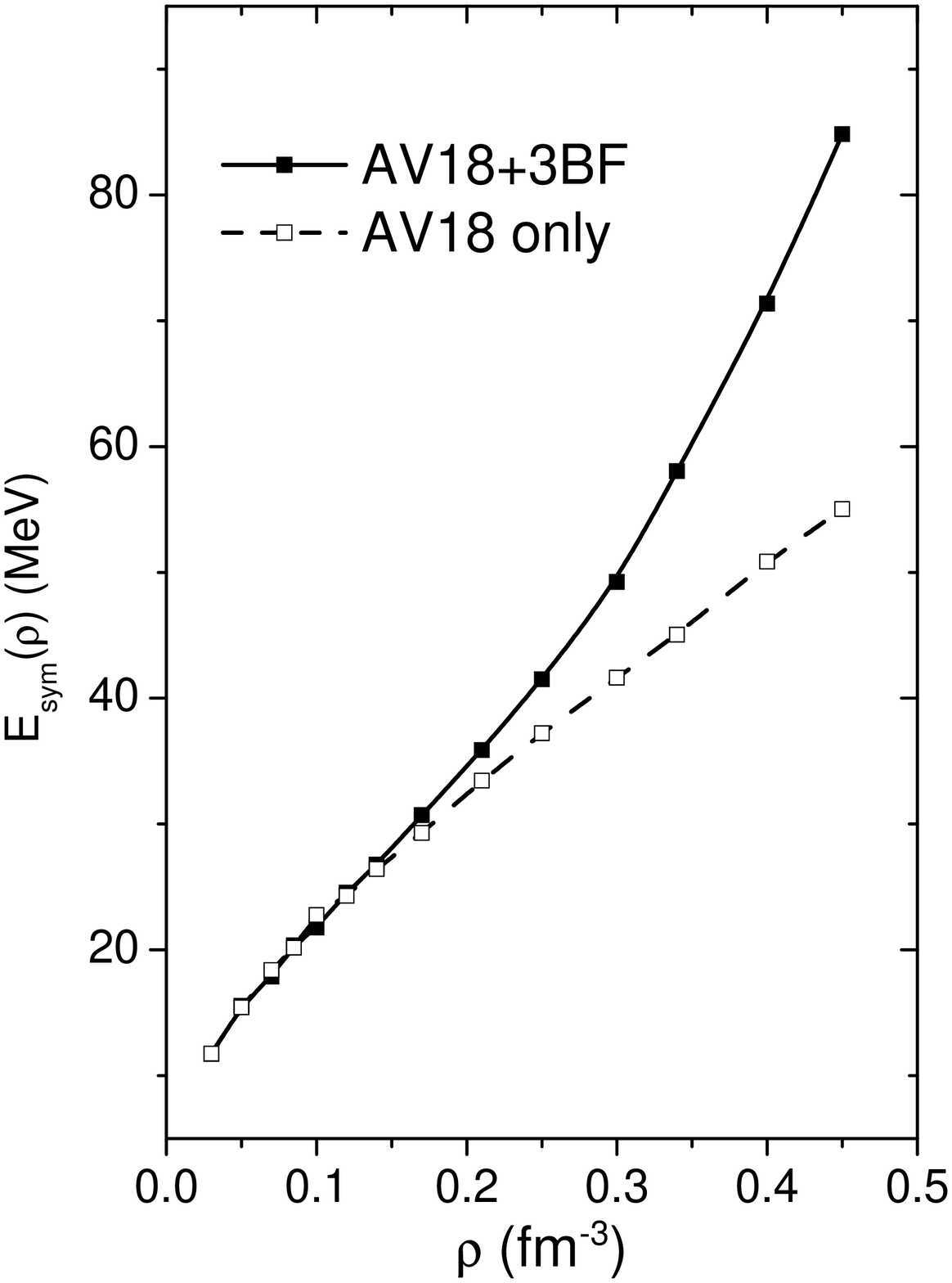}
\vspace{-7mm}
\caption{Isospin symmetry energy from BHF approximation with and without 3BF.}
\end{minipage}
\hspace{7mm}
\begin{minipage}[t]{74mm}
\includegraphics[angle=0,width=65mm,height=7cm]{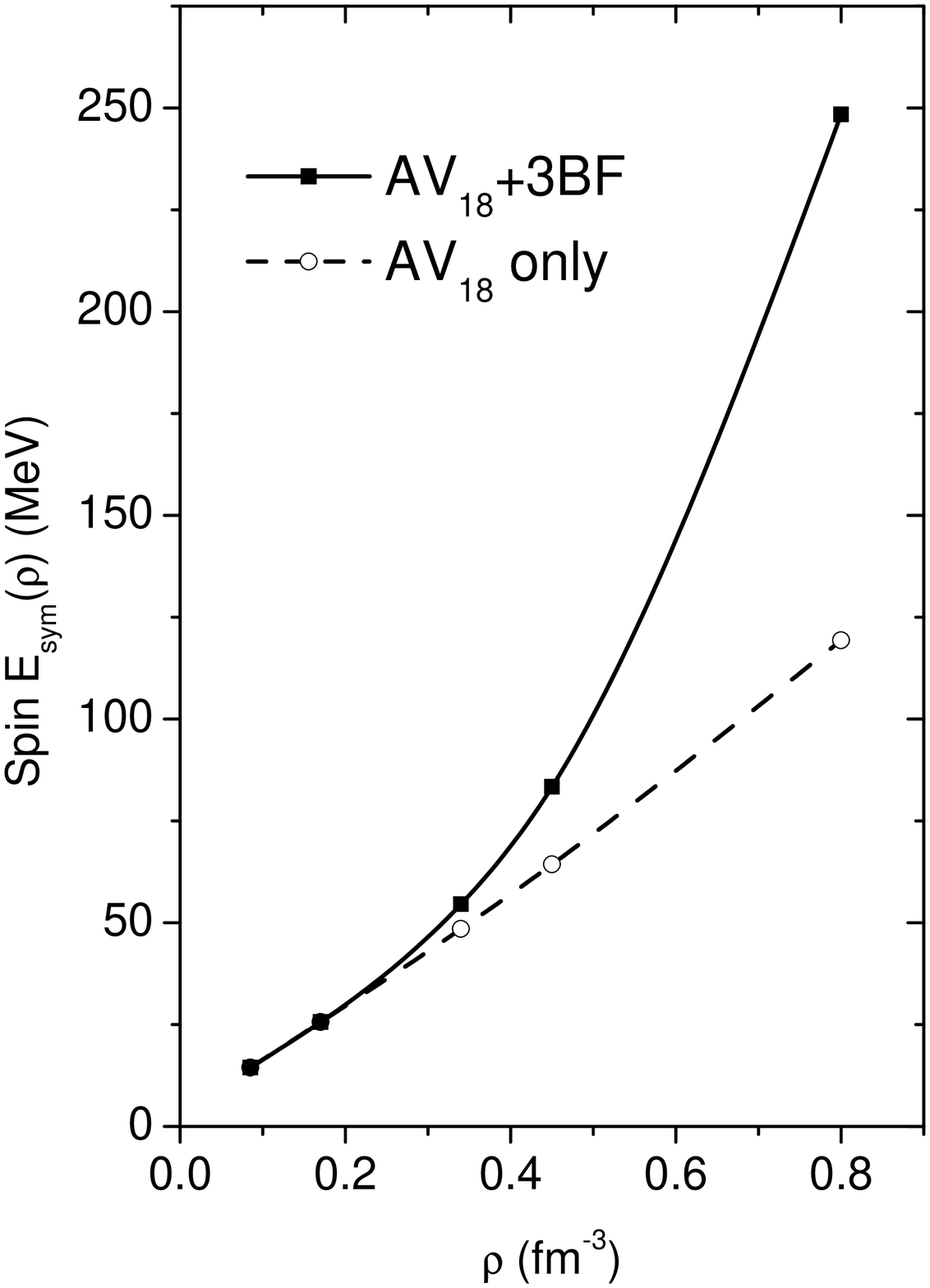}
\vspace{-7mm}
\caption{Spin symmetry energy from BHF approximation with and without 3BF.}
\end{minipage}
\end{figure}
  
\section{LANDAU PARAMETERS FROM THE BRUECKNER THEORY} 

The Brueckner theory basically provides us with the $G$-matrix, which is 
nothing else than the in-medium NN interaction. The latter is not
suitable for calculations, but it can be cast in the form of 
effective interaction within the quasiparticle description of Fermi systems
\cite{BROW,BAL}. Alternatively, for the present purpose  one first determines 
from the EOS of nuclear matter fundamental equilibrium properties such
as the compression modulus, symmetry energy and spin susceptibility and then
calculates the corresponding  Landau parameters.  
\begin{figure}
\begin{minipage}[t]{74mm}
\includegraphics[angle=0,width=73mm]{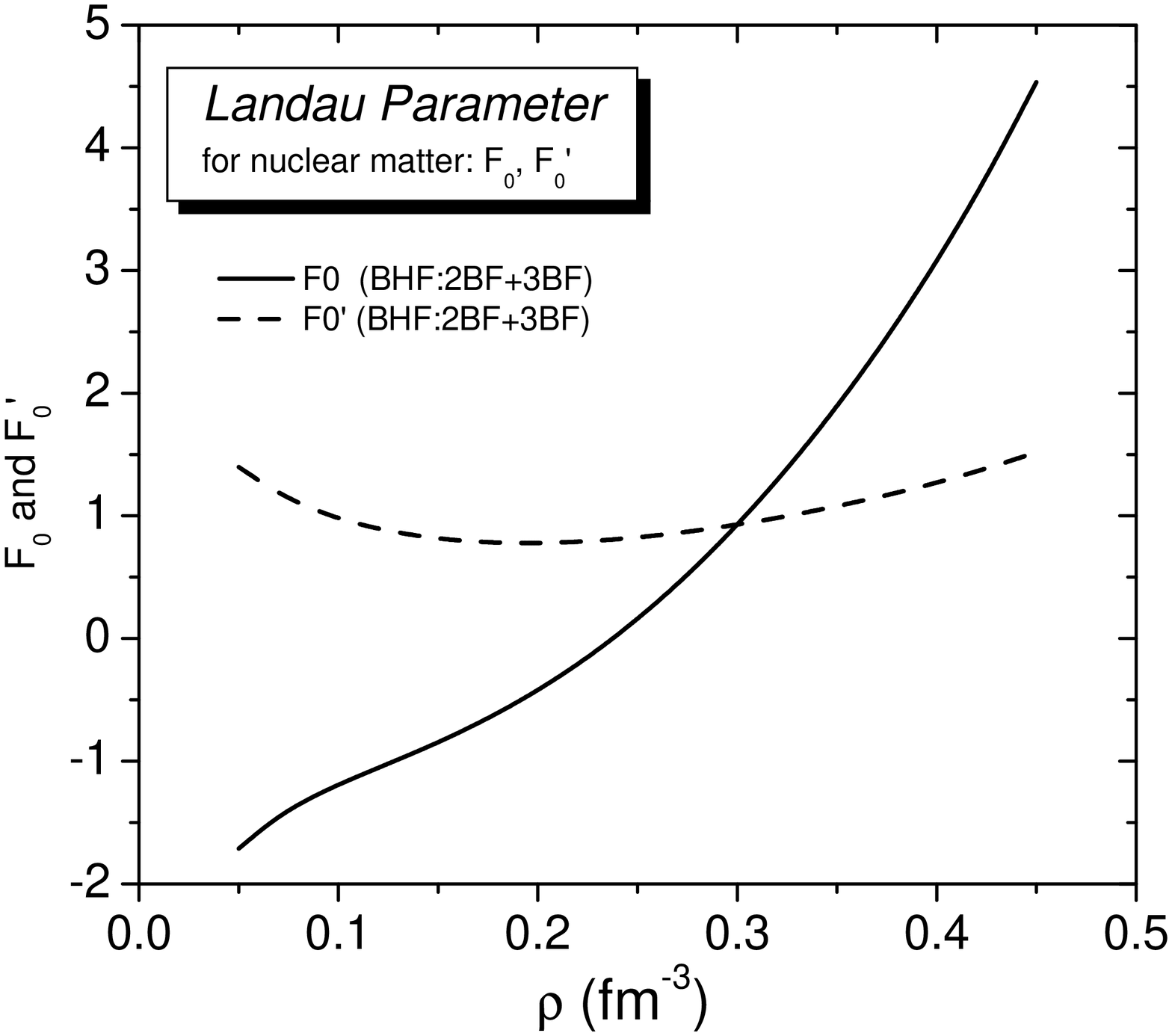}
\vspace{-7mm}
\caption{Landau parameters $F_0$ and $F_0$' in nuclear matter.}
\end{minipage}
\hspace{5mm}
\begin{minipage}[t]{74mm}
\includegraphics[angle=0,width=73mm,height=64mm]{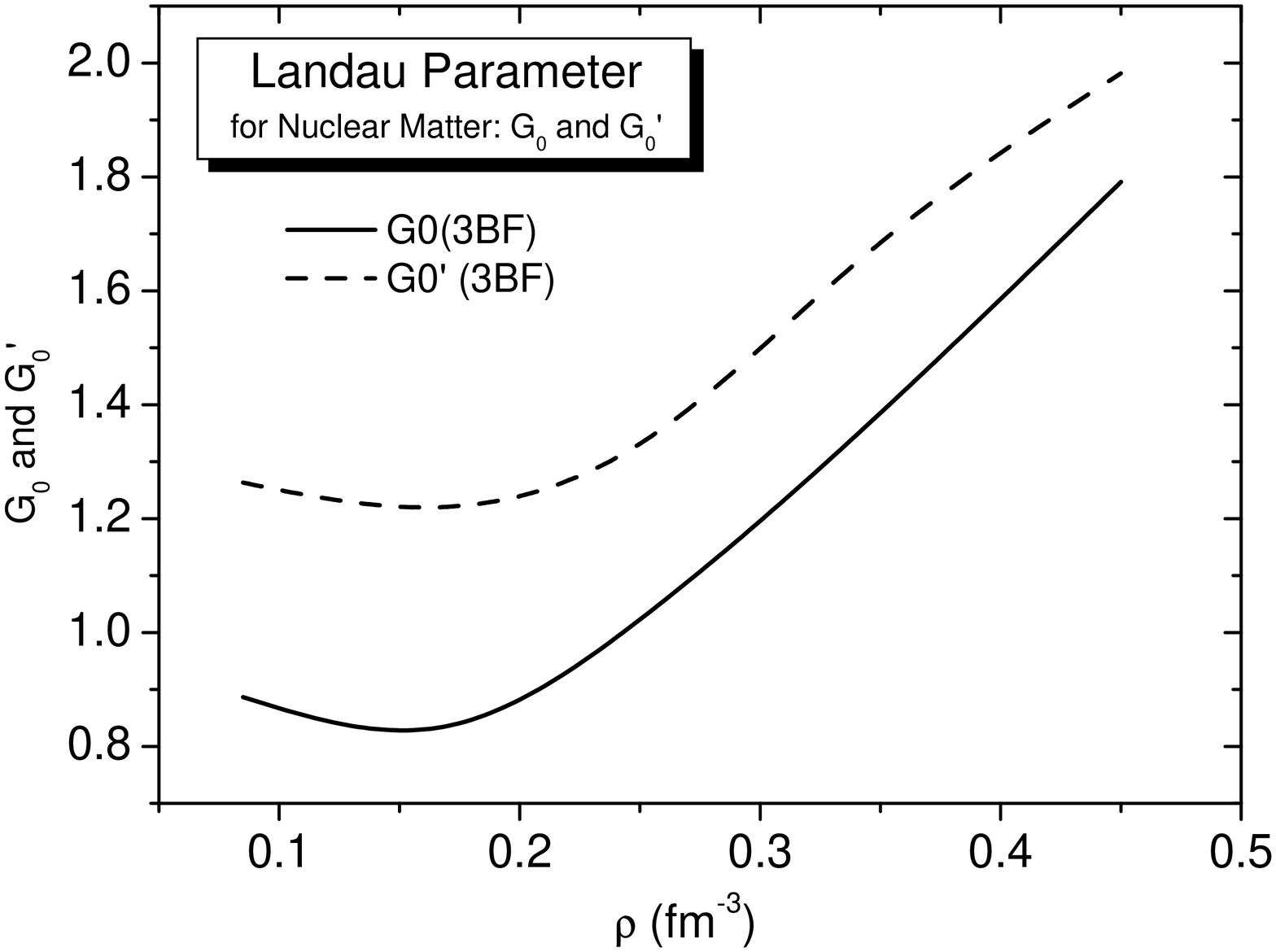}
\vspace{-7mm}
\caption{Landau parameters $G_0$ and $G_0$' in nuclear matter.}
\end{minipage}
\end{figure}

In the Landau theory of Fermi liquids one writes the  
particle-hole residual interaction as
\begin{equation}
N(0){\cal V}_{NN} = F + F' \vec\tau\cdot\vec\tau' + G \vec\sigma\cdot\vec\sigma'
+ G'\vec\tau\cdot\vec\tau'\vec\sigma\cdot\vec\sigma',
\end{equation}
where $F,F',G,G'$ are the Landau parameters. Each of them is expanded in
a Legendre series, for instance $F=\sum F_l P_l(cos\theta)$ where $\theta
$ is the Landau angle. The physical quantities related to the static response 
functions can be expressed in a simple way in terms of the Landau parameters.     
The response to density fluctuation, i.e. the compression modulus is directly
related to $F_0$; the response to charge variation, i.e. the symmetry energy, is
related to $F'_0$; the response to spin variation, i.e. the magnetic susceptibility, 
is related to $G_0$; finally, the response to spin-isospin variation, which is the 
isospin susceptibility, is related to $G'_0$. 

From the Brueckner calculations we found that the energy per nucleon of nuclear matter 
as a function of $\delta_n$ and $\delta_p$ 
(Eq.~(1) for neutron and proton, separately) can be fairly approximated by the quadratic form
\begin{equation}
E_A(\delta_n,\delta_p)-E_A(0,0)=\Delta_{nn}\delta_n^2 + \Delta_{pp}\delta_p^2+
2 \Delta_{np}\delta_n \delta_p .
\end{equation} 
This equation is the basis of our calculation of the zero-order Landau parameters,
which are plotted in Fig. 4 and Fig. 5. Notice that $F_0$ becomes less than -1 
 (negative compression modulus)  at low density which corresponds to the unphysical 
 spinodal region of the EOS, signaled by the multifragmentation events of heavy ion collisions. 

\begin{table}
\caption{Landau parameters of neutron matter from Brueckner (BRU) calculations with 
Argonne $V_{18 }$ and microscopic 3BF described in the text. Compression modulus and 
magnetic susceptibility are also reported in comparison with Monte Carlo (MC) 
calculations using $V_{18 }$ plus phenomenological UIX 3BF \cite{FANT} and Hartree-Fock 
calculations (SLy230b) with  Skyrme-like force \cite{CHAB}.}
\renewcommand{\tabcolsep}{4mm} 
\renewcommand{\arraystretch}{1.1} 
\begin{tabular}{l|ll|ccc|ccc}
\hline
& $F_{0}$ & $G_{0}$ & \multicolumn{3}{|c}{$K/K_{F}$} & \multicolumn{3}{|c}{$%
\chi /\chi _{F}$} \\ 
\multicolumn{1}{c|}{$\rho $(fm$^{-3})$} & \multicolumn{2}{|c|}{BRU} & BRU & 
MC & SLy & BRU & MC & SLy \\ \hline
\multicolumn{1}{c|}{0.20} & \multicolumn{1}{|c}{0.615} & \multicolumn{1}{c|}{
1.170} & 1.84 & 2.13 & 1.45 & 0.41 & 0.37 & 0.32 \\ 
\multicolumn{1}{c|}{0.32} & \multicolumn{1}{|c}{3.395} & \multicolumn{1}{c|}{
1.461} & 5.24 & 4.76 & 4.32 & 0.37 & 0.33 & 0.22 \\ \hline
\end{tabular}
\end{table}

\section{NEUTRINO MEAN FREE PATH}

Let us consider the propagation of neutrinos in pure neutron matter.
The neutrino mean free path $\lambda$ is derived from the transport equation.
One obtains
\begin{equation}
\frac{c}{\lambda_{\vec k}} = \sum W_{fi}(k-k')(1-n_{\vec k'})+
W_{fi}(k'-k) n_{\vec k'},
\end{equation} 
where $n_{\vec k}$ is the occupation number of neutrinos and $W_{fi}$ is
the neutrino-neutron scattering probability via the weak neutral current\cite{IWA}. 
The medium effects are incorporated into $W_{fi}$ via
the dynamical form factors, 
\begin{figure}[htb]
\vspace{-2mm}
\centering{\includegraphics[width=12cm,height=6cm]{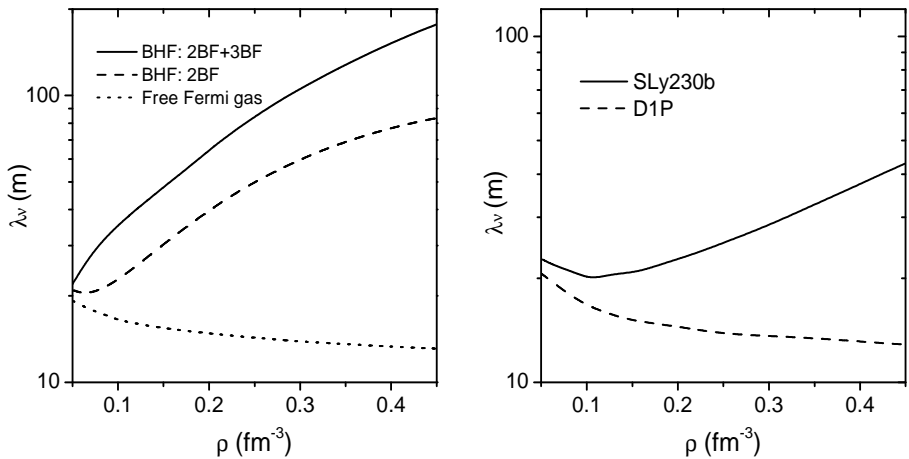}}
\vspace{-9mm}
\caption{RPA neutrino mean free path vs density at temperature $T=10$ MeV, and
neutrino energy $E_{\nu}=30$ MeV (left) from Brueckner calculations and comparison 
with Hartree-Fock results calculated with phenomenological Skyrme and Gogny forces (right).}
\vspace{-1mm}
\end{figure}
which, in turn, are related to the imaginary part of the dynamical response 
function. In the low-frequency limit 
${\omega}/({kv_F}) \ll 1$ the latter is expressed in a simple way 
in terms of the Landau parameters\cite{IWA}, which is the approximation adopted 
in the present calculation. The Landau parameters were extracted 
from the Brueckner prediction of the EOS of neutron matter. They are 
listed in Table 1 in comparison with other predictions.
In Fig.~6  (left panel) our prediction for $\lambda_\nu$ in interacting neutron
matter is reported as a function of the density in comparison with 
that in a free neutron
gas. One can see that the large medium effects are due to the large value taken by both Landau
parameters $F_0$ and $G_0$ in the interacting system.
For densities in the range of neutron-star density $\lambda_\nu$ is by one order 
of magnitude larger than in a free Fermi gas.
In the right panel of Fig. 6 the Hartree-Fock predictions from two Skyrme forces
and Gogny force taken from Ref.\cite{MARG} are reported for comparison. 
 In our case $\lambda_\nu$ is
increasing much faster with density due to the more repulsive character
of our 3BF.    
Fig. 7 shows the dependence of $\lambda_\nu$ on the temperature T. The expected  
T dependence of the Landau parameters is neglected.   
The relative weight of each elementary process giving rise to the neutrino 
propagation is plotted in Fig.~8. The role played by the Landau zero mode 
appearing as a singularity of the response function, is negligible.

\begin{figure}[htb]
\vspace{-2mm}
\begin{minipage}[t]{74mm}
\includegraphics[width=73mm,height=6cm]{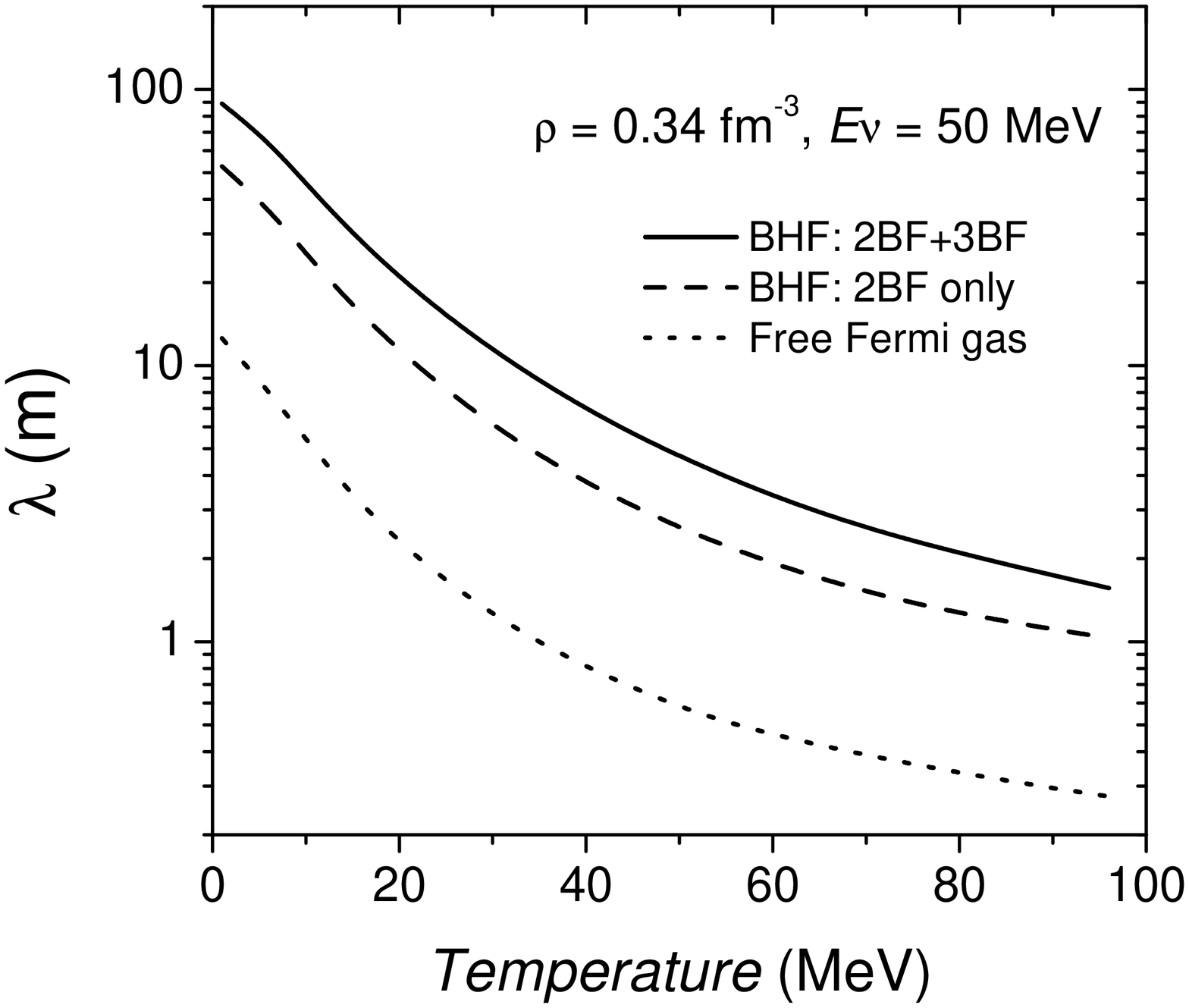}
\vspace{-9mm}
\caption{Neutrino mean free path vs. temperature obtained by BHF calculations 
with (solid) and without (dashes) TBF, in comparison with the free Fermi gas model (dotted line). }
\end{minipage}
\hspace{7mm}
\begin{minipage}[t]{74mm}
\includegraphics[width=73mm,height=6cm]{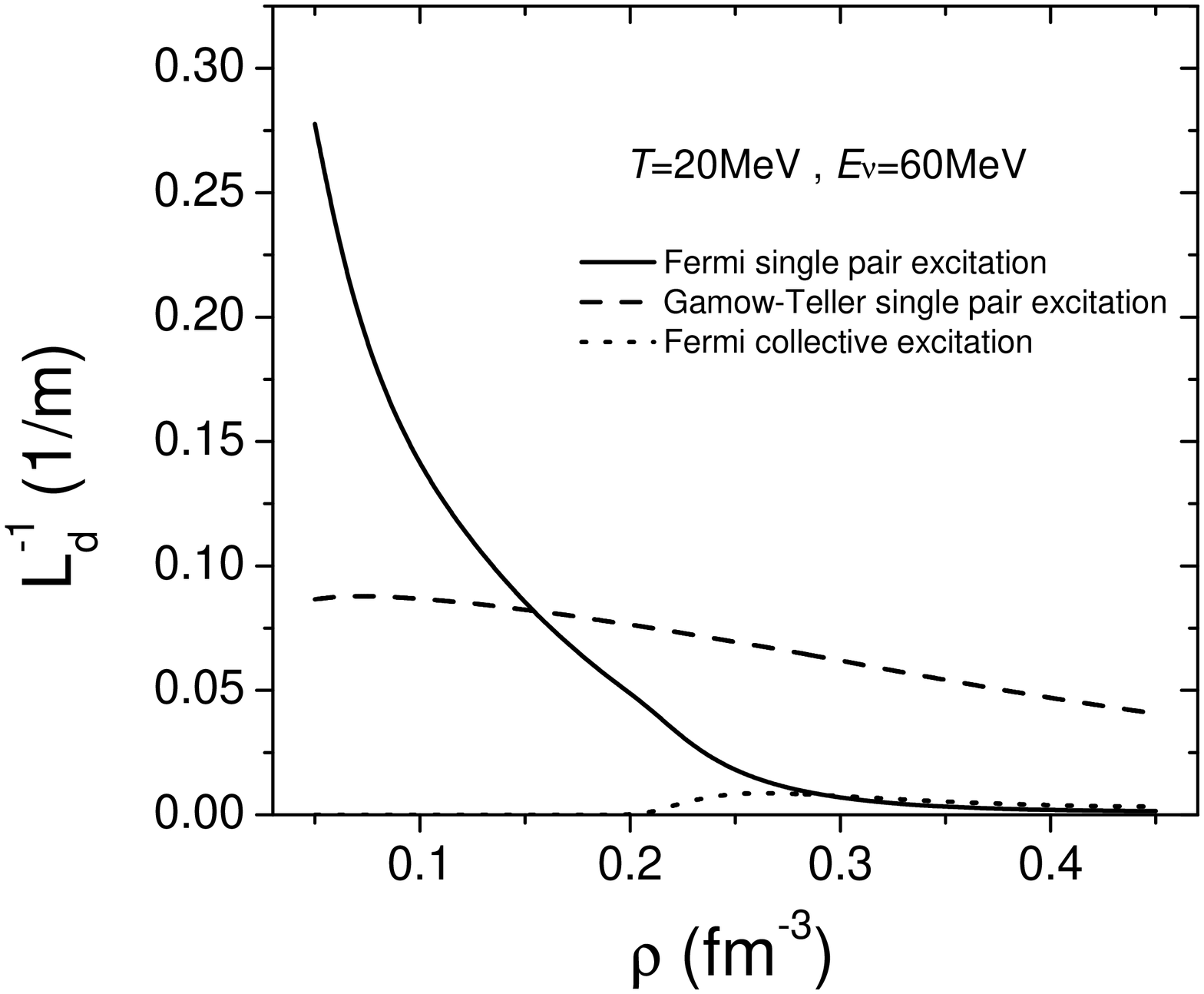}
\vspace{-9mm}
\caption{Partial contributions to the neutrino mean free path due to different excitation modes.}
\end{minipage}
\vspace{-2mm}
\end{figure} 

In summary, we have presented  a prediction of  mean free path of neutrinos 
in neutron matter. Since the strong nuclear correlations enhance the 
propagation of neutrino one may predict a faster cooling of a neutron stars
via neutrino escape than in a degenerate neutron gas. The neutron star
temperature plays also important role in the neutrino propagation. Due to
the large value of $T$ involved in proton-neutron stars, the present predictions 
should be improved by introducing $T$ dependent Landau parameters.

This work has been supported in part by the Chinese Academy of Science,
within the {\it one Hundred Person Project}, the Knowledge
Innovation Project of CAS under No. KJCX2-SW-N02, and the Important
Pre-research Project of the  Ministry of Science and technology under
No.2002CCB00200, China.

\end{document}